\documentclass[10pt,preprint,aps]{revtex4-1}
\usepackage[latin9]{inputenc}
\usepackage{color}
\usepackage{textcomp}
\usepackage{amsmath}
\usepackage{amssymb}
\usepackage{graphicx}
\usepackage{subscript}
\usepackage[unicode=true,
 bookmarks=true,bookmarksnumbered=false,bookmarksopen=false,
 breaklinks=false,pdfborder={0 0 1},backref=false,colorlinks=false]
 {hyperref}
\hypersetup{pdftitle={quantum state preparation}}



\begin{document}

\title{\textcolor{black}{Rotational spectroscopy of cold, trapped molecular
ions} \textcolor{black}{in the Lamb-Dicke regime }}

\author{S. Alighanbari}

\address{Institut für Experimentalphysik, Heinrich-Heine-Universität Düsseldorf,
40225 Düsseldorf, Germany}

\author{M.~G. Hansen}

\address{Institut für Experimentalphysik, Heinrich-Heine-Universität Düsseldorf,
40225 Düsseldorf, Germany}

\author{V.~I. Korobov}

\address{Bogoliubov Laboratory of Theoretical Physics, Joint Institute for
Nuclear Research, 141980 Dubna, Russia}

\author{S. Schiller{*}}

\address{Institut für Experimentalphysik, Heinrich-Heine-Universität Düsseldorf,
40225 Düsseldorf, Germany}
\begin{abstract}
\textbf{\textcolor{green}{}}\textbf{ Sympathetic cooling of trapped
ions has been established as a powerful technique for manipulation
of non-laser-coolable ions \cite{Raizen1992,Waki1992,Bowe1999,Barrett2003}.
For molecular ions, it promises vastly enhanced spectroscopic resolution
and accuracy. However, this potential remains untapped so far, with
the best resolution achieved being not better than $5\times10^{-8}$
fractionally, due to residual Doppler broadening being present in
ion clusters even at the lowest achievable translational temperatures
\cite{Bressel2012}. Here we introduce a general and accessible approach
that enables Doppler-free rotational spectroscopy. It makes use of
the strong radial spatial confinement of molecular ions when trapped
and crystallized in a linear quadrupole trap, providing the Lamb-Dicke
regime for rotational transitions. We achieve a line width of $1\times10^{-9}$
fractionally  and 1.3~kHz absolute, an improvement by $50$ and
nearly $3\times10^{3}$, respectively, over other methods. The systematic
uncertainty is $2.5\times10^{-10}$.  As an application, we demonstrate
the most precise test of }\textbf{\textit{ab initio}}\textbf{ molecular
theory and the most precise (1.3~ppb) spectroscopic determination
of the proton mass. The results represent the long overdue extension
of Doppler-free microwave spectroscopy of laser-cooled atomic ion
clusters \cite{Berkeland1998} to higher spectroscopy frequencies
and to molecules. This approach enables a vast range of high-precision
measurements on molecules, both on rotational and, as we project,
vibrational transitions. }
\end{abstract}
\maketitle
Rotational spectroscopy of gas-phase molecules is a time-honored technique
which has been fundamental for developing our knowledge of molecular
structure \cite{Townes1975}. Its range continues to be extended,
both in laboratory, in astronomical observatories and in space instruments,
for example to cold molecules \cite{Jusko2014}. Nevertheless, rotational
spectroscopy has, in the vast majority of cases, not been able to
achieve ultra-high spectroscopic resolution and accuracy.  Enabling
this would open up numerous opportunities for studies in molecular
physics and in fundamental physics, such as tests of molecular quantum
theory, measurement of magnetic and optical susceptibilities, investigation
of collision interactions, tests of the time-independence of particle
masses \cite{Schiller2005,Shelkovnikov2008,Uzan2011,Godun2014,Huntemann2014},
and the measurement of fundamental constants \cite{Roth2008b,Koelemeij2007,Bressel2012,Biesheuvel2016}. 

In conventional (linear) rotational spectroscopy resolution can be
improved by cooling of the molecules to a cryogenic translational
temperature $T$, but the gains possible with thermal cooling methods
($T\gtrsim10\,{\rm K}$) are modest \cite{Jusko2014}, due to the
$\sqrt{T}$ - dependence of the Doppler line width. For untrapped,
neutral molecules, a leap in resolution and accuracy in rotational
spectroscopy was achieved with the introduction of Lamb-dip (saturation)
spectroscopy \cite{Winton1970,Cazzoli1990}. It allowed improving
the fractional line resolution by approximately a factor 20 - 30 beyond
the Doppler broadening, to $5\times10^{-8}$ \cite{Winton1970}.
This level has, however, not improved for nearly 50~years \cite{Winnewisser1997,Cazzoli2013},
since it is limited by time-of-flight broadening. In the context
of molecular ions, ion trapping combined with sympathetic cooling
by Doppler-laser-cooled atomic ions to the ``crystallized'' cluster
state provides a well-tested approach for further reducing $T$ to
the 10~mK level, with concomitant reduction of Doppler line width
by approximately a factor 30 \cite{Bressel2012}. 

Here, for the first time, we not only realize this improvement for
rotational spectroscopy but also achieve Doppler-free spectral resolution,
by reaching the rotational Lamb-Dicke regime (LDR). The LDR is defined
by $\delta x<\lambda_{{\rm rot}}/2\pi$, where $\delta x$ is the
range of the ions' motion along the spectroscopy beam direction ${\bf \hat{k}\,}||\,x$
\cite{Dicke1953}. In this work, the motion ranges are thermal in
origin. In prolate ion clusters of appropriate size, the ranges of
motion orthogonal to the clusters' long axes ($z$) are $\delta x,\,\delta y<20\,\mu{\rm m}$.
Directing ${\bf \hat{k}}$ orthogonal to $z$ satisfies the LDR condition
for rotational transition wavelengths, with typical values $\lambda_{{\rm rot}}\simeq0.2\,-\,2\,$mm.
Ultra-high fractional and absolute frequency resolution are thereby
enabled. In contrast, vibrational spectroscopy, where wavelengths
are $\lambda_{{\rm vib}}<8\,\mu\text{m}$, yields lines exhibiting
the classic Doppler width \cite{Bressel2012,Germann2014}, often further
complicated by unresolved hyperfine structure \cite{Koelemeij2007,Biesheuvel2016}.
  The present methods does not require complex techniques such as
single-ion manipulation, ground-state cooling, and quantum spectroscopy
\cite{Wolf2016a,Chou2017} to capitalize on the advantage of the
LDR regime. The performance improvement in terms of fractional resolution
is a factor 50 compared to both previous trapped molecular ion ensemble
spectroscopy \cite{Bressel2012,Germann2014} and\textbf{} to the
highest resolution rotational spectroscopy of neutral molecules reported
so far \cite{Cazzoli2013}, to the best of our knowledge. 

In order to perform a stringent test of the new method, we choose
the polar molecule with the  smallest fundamental rotational transition
wavelength $\lambda_{{\rm rot,min}}\simeq228\,{\rm \text{\ensuremath{\mu}m}}$
($f_{{\rm rot,max}}\simeq1.3\,{\rm THz}$): ${\rm HD}^{+}$. In addition,
the feasibility of \textit{ab initio} calculation of $f_{{\rm rot}}$
for HD\textsuperscript{+} and of its sensitivities to external fields
allows testing the spectroscopic \textit{accuracy} of the method.

\begin{figure}[t]
\begin{centering}
\includegraphics[width=1\columnwidth]{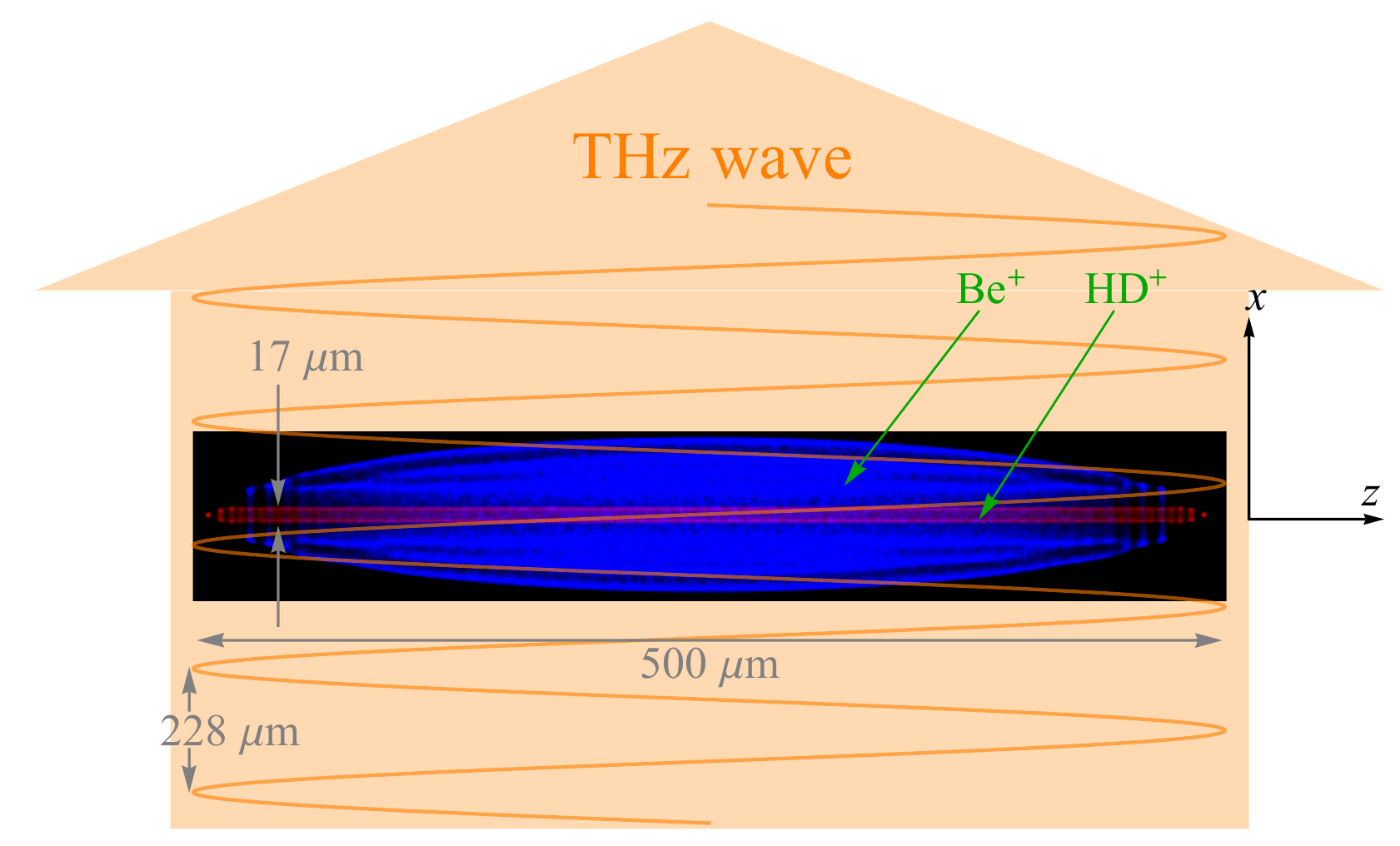}
\par\end{centering}
\caption{\label{fig:Basic principle} Principle of the Lamb-Dicke rotational
spectroscopy of sympathetically cooled molecular ions. The ion cluster
is prolate, and the sympathetically cooled ions exhibit a relatively
small motional range in the directions $x,\,y$ perpendicular to the
trap axis $z$. The spectroscopy radiation propagates perpendicular
to $z$. The ion cluster image is a time average of ion trajectories
obtained from an MD simulation of an ensemble of $N=200\,\,{\rm HD}^{+}$
ions and $N_{{\rm Be}^{+}}=2000$ atomic ions (see Methods). Ion
clusters generated in the experiment are similar to the one shown
here. }
\end{figure}

The basic concept is depicted in Fig.~\ref{fig:Basic principle}.
In a prolate ion cluster of appropriate size, trapped in a linear
quadrupole trap (trap axis is along $z$), the sympathetically cooled
ions arrange approximately as a narrow cylinder aligned along the
trap axis. Maximum ion radial distances from the axis are significantly
smaller than the cylinder length. Molecular dynamics simulations for
a cluster containing $N=200$ HD\textsuperscript{+} ions (Cluster
1, see Methods) indicate r.m.s. position variations (corresponding
approximately to the range $\delta x/2$) $\Delta x=\Delta y\simeq8.4\,\mu\text{m}$
at $T=12$~mK, increasing slightly to $9.0\,\mu$m at 67~mK. These
values are significantly smaller than $\lambda_{{\rm rot}}/2\pi\simeq36\,\mu\text{m}$,
indicating the LDRif ${\bf \hat{k}}$ is chosen perpendicular to
the trap axis.

In the LDR, the absence of Doppler broadening and, because of ion
confinement, also of transit-time broadening, puts into evidence other
broadening mechanisms. Among these, pressure (collision) broadening,
can be suppressed by operating under a sufficiently small residual
background neutral gas pressure. Collisions between the trapped ions
themselves do not lead to appreciable frequency shifts, because minimum
approach distances are of order$10\,\mu$m. 

A major broadening effect can then be power (saturation) broadening
by the spectroscopy wave. The corresponding line width is $\Delta\nu_{{\rm pb}}=\sqrt{2}\Omega_{{\rm R}}/2\pi$,
with the Rabi angular frequency $\Omega_{{\rm R}}=\mu_{eg}E/\hbar$,
the transition dipole moment $\mu_{eg}$, and the electric field amplitude
$E$. A beam power $P=\frac{1}{2}\epsilon_{0}c\,E^{2}A=1\,{\rm pW}$,
and a beam cross sectional area $A=\pi\times1\,\text{mm}^{2}$ (limited
from above by the distance between the trap electrodes, and by diffraction),
yield $\Delta\nu_{{\rm pb}}=43$~Hz, assuming $\mu_{eg}=0.15\,e\,a_{0}$
 (as for a ``strong'' spin component of the HD\textsuperscript{+}
fundamental rotational transition). Thus, extremely low, pW-level
power must be used on ``strong'' transitions if the line broadening
is to be reduced below the $10^{-10}$ level, even for $f_{{\rm rot}}$
as high as $10^{12}$~Hz. It is advantageous that such low power
levels strongly simplify the technological requirements on the source.

Ultra-narrow transitions may be challenging to find. Theoretical predictions
and results from e.g. lower resolution rovibrational spectroscopy
may be helpful or necessary to reduce the spectral range of the search.
Also, performing the spectroscopy initially at high intensity leads
to broad spectral lines, which are easier to find. Subsequently, the
intensity is progressively reduced to increase spectral resolution
and accuracy. 

In order to access experimentally the smallest possible line widths,
it is necessary to lift as far as possible the Zeeman degeneracy of
transitions by applying a magnetic field. Excitation of an individual
Zeeman component implies excitation of the molecules populating a
single quantum state. Techniques for increasing the population in
this state and thus the signal may therefore be required. Here, we
use rotational laser cooling (see Methods).

The potential of rotational spectroscopy in the LDR can only be harnessed
fully if the microwave source has excellent spectral purity (small
line width) and high absolute frequency stability. For this work,
we have implemented a ``frequency chain'', where $f_{{\rm rot}}\simeq1.3\,{\rm THz}$
is referenced to a hydrogen maser (see Methods). 

\begin{figure}[t]
\centering{}\includegraphics[width=0.9\columnwidth]{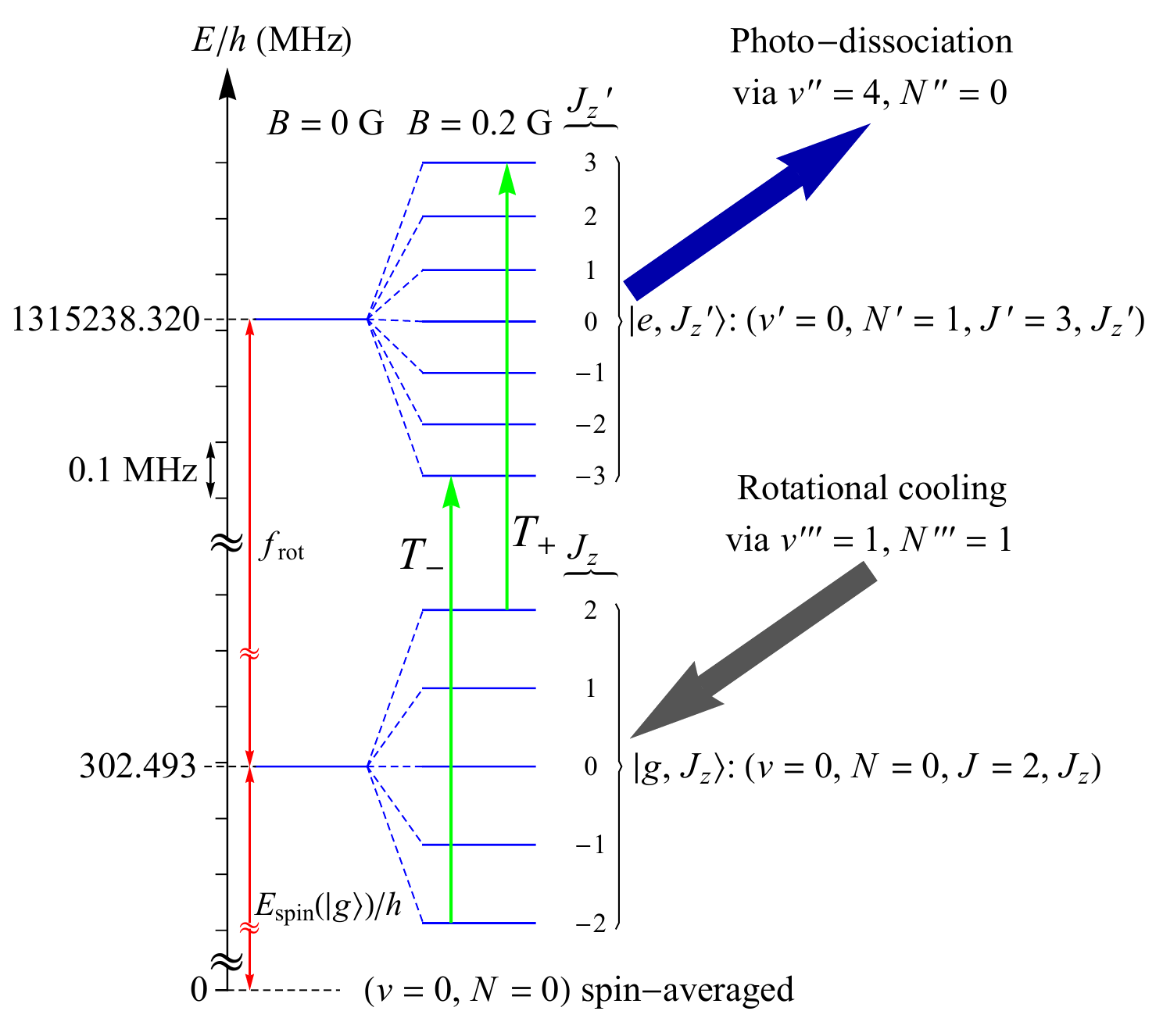}\caption{\label{fig:Energy diagram-1}Simplified diagram of relevant energy
levels of HD\protect\textsuperscript{+} in the ground vibrational
level $v=0$. We show only one spin state, $J=2$, of the ground rotational
level $N=0$, denoted by $|g\rangle$, and one spin state, $J'=3$,
of the first excited rotational level $N'=1$, denoted by $|e\rangle$.
The green arrows show the transitions addressed in this work, exhibiting
a linear Zeeman splitting. The population of the $N=0$~-~spin
states is enhanced by rotational cooling (thick gray arrow), and the
population in the $N'=1$~-~states is photo-dissociated by two lasers
(blue arrow). The zero of the energy scale corresponds to the energy
of the $N=0$ - level in absence of the effective spin Hamiltonian
interaction of ref.~\cite{Bakalov2006}.}
\end{figure}

We address one particular spin component of the rotational transition,
$|g\rangle=(v=0,\,N=0,\,J=2)\xrightarrow{1.3\:\text{THz}}$ $|e\rangle=(v'=0,\,N'=1,\,J'=3)$,
see Fig.~\ref{fig:Energy diagram-1}. $v,\,N,\,J$ denote vibrational,
rotational and total angular momentum quantum number, respectively.
Spectroscopy is performed by $1+1'+1''$ resonance-enhanced multi-photon
dissociation (REMPD), where the upper spectroscopy level undergoes
$|e\rangle\xrightarrow{1.42\:\mu\text{m}}$ $(v''=4,\,N''=0,\,J'''=2)$
$\xrightarrow{266\:\text{nm}}\,{\rm H+D}^{+}$. The reduction of
the number of trapped HD\textsuperscript{+} upon dissociation is
measured using secular excitation, and represents the spectroscopy
signal. Repeated HD\textsuperscript{+} trap loading and spectroscopy
cycles are performed and the signals averaged so as to increase the
signal-to-noise ratio. Typical ion clusters used for spectroscopy
are similar to the one shown in Fig.~\ref{fig:Basic principle},
having $T\simeq10\,$mK.

\begin{figure}[t]
\begin{centering}
\includegraphics[width=0.9\columnwidth]{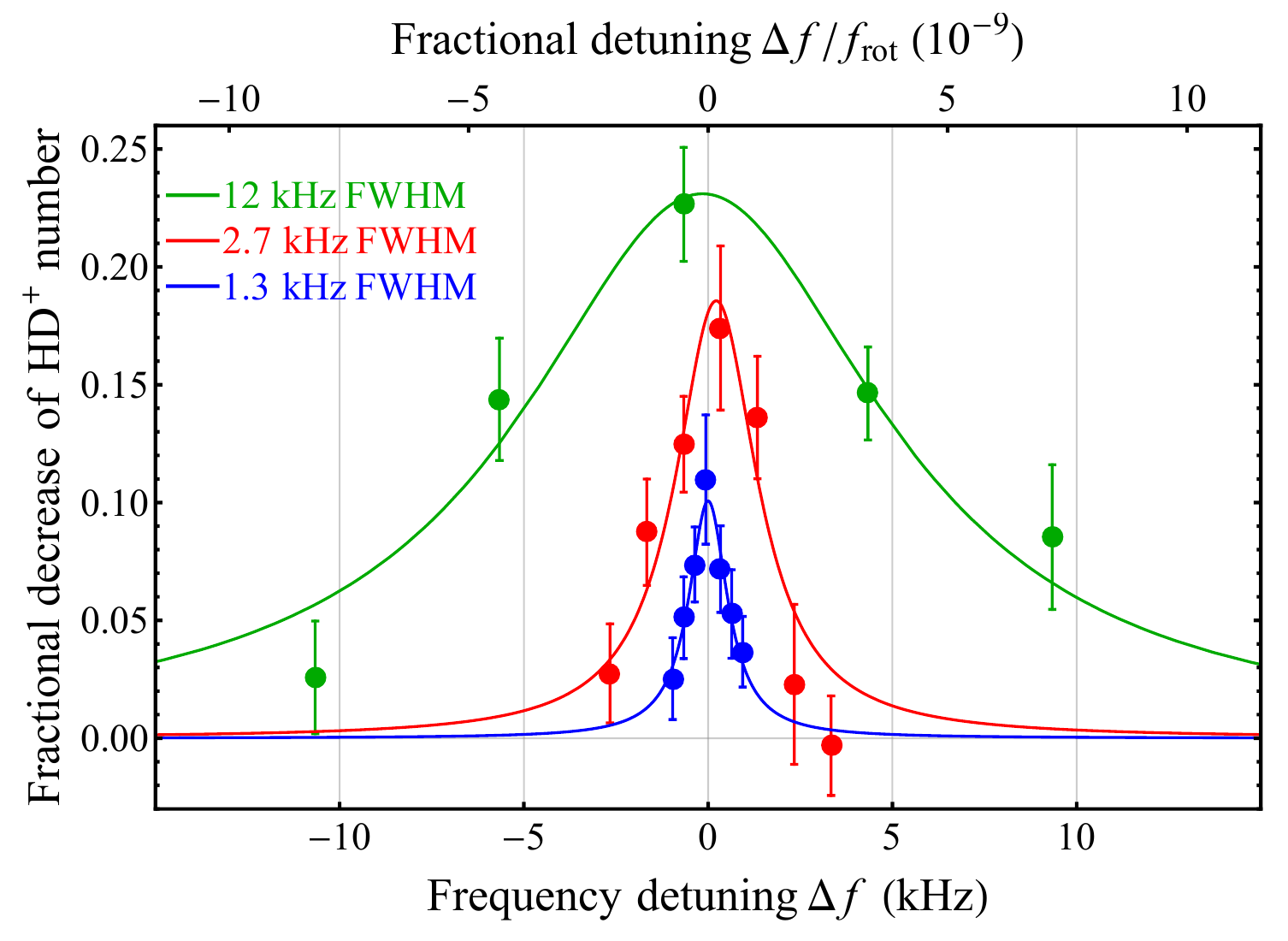}
\par\end{centering}
\caption{\label{fig:Power-broadening-1}Spectroscopy of one hyperfine component
of the fundamental rotational transition of HD\protect\textsuperscript{+}.
Shown is the unresolved transition pair T\protect\textsubscript{$\pm$}
between the stretched states, $(J=2,\,J_{z}=\pm J)\rightarrow(J'=3,\,J_{z}'=\pm J')$,
at different power levels of the THz wave. The magnetic field is $B\simeq0.4$~G. The
lines are Lorentzian fits. Zero detuning corresponds to the center
frequency of the blue fit.}
\end{figure}

Fig.~\ref{fig:Power-broadening-1} shows the measured spectrum $|g\rangle\rightarrow|e\rangle$
in the neighborhood of the theoretically predicted frequency (see
Methods)

\begin{equation}
f_{{\rm rot,theory}}=1\,314\,935.827\,3(10)~{\rm MHz\,.}\label{eq:f_rot,theory}
\end{equation}
  At comparatively high intensity (approximately $0.1\,\mu\text{W}/\text{mm}^{2}$,
green points) the line width is 12~kHz, whereas the calculated Doppler
line width is 54~kHz. By reducing the source power successively the
line width decreases to 1.3~kHz (blue points). This clearly evidences
power broadening. The latter line width is mostly due to residual
power broadening. In order to obtain sufficient signal strength, we
did not lift the Zeeman degeneracy completely and the spectrum contains
the unresolved superposition of the Zeeman transition pairs T\textsubscript{+}:~$J_{z}=2\rightarrow J_{z}'=3$
and ${\rm {\rm T_{-}}:\,}J_{z}=-2\rightarrow J_{z}'=-3$ ($J_{z}$
is the projection of the total angular momentum on the quantization
axis). . The closest other Zeeman component has a detuning of $-13$~kHz
or more.

The shift of the transitions ${\rm T}_{\pm}$ with magnetic field
$B$ has been theoretically calculated to be strictly linear in $B$,
$\Delta f({\rm T_{\pm},\,}B)=\pm E_{10}(v=0,\,N=1)\,B=\mp0.56\,\text{kHz}\,(B/1\,{\rm G})$
(\cite{Bakalov2011} and Methods). The mean frequency is therefore
free of any Zeeman shift. The magnetic field is $B=0.40(6)\,{\rm G}$,
measured with an anisotropic magnetoresistive probe and by radio-frequency
spectroscopy of the Be\textsuperscript{+} ions using the method of
ref.~\cite{Shen2014}. 

We obtain the rotational frequency 

\begin{equation}
f_{{\rm rot,exp}}=1\,314\,935.828\,0(6)~{\rm MHz}\,.
\end{equation}

 The uncertainty $(4\times10^{-10})$ results from the experimental
resolution and from the Zeeman pair asymmetry uncertainty $(2.5\times10^{-10})$.
Other perturbations of the transition frequency are compatible with
zero within an estimated uncertainty $3\times10^{-11}$ (see Methods).

The value $f_{{\rm rot,exp}}$ is in agreement with the \textit{ab
initio} prediction $f_{{\rm rot,theory}}$ within the combined error,
1.1~kHz. We thus confirm the accuracy of the molecular calculations,
including its QED contributions, at  the level of 1.1~kHz $(9\times10^{-10})$,
limited by theory. This represents the most precise test of a molecular
physics prediction yet. 

The \textit{ab initio} value $f_{{\rm rot,theory}}$ is based on the
CODATA2014 values of the fundamental constants \cite{Mohr2016}, in
particular on the proton mass $m_{p}$. Among the stable fundamental
particles of atomic physics, this is the particle that currently has
the largest fractional uncertainty (0.09~ppb). We can obtain a value
for $m_{p}$ from the present experiment by treating it as a fit parameter,
while taking the other constants, $m_{d}$, $m_{e}$, $R_{\infty}$,
$\alpha$, and their uncertainties from CODATA2014. The spectroscopically
determined value then is 

\begin{equation}
m_{p}=1.007\,276\,466\,9(13)\,{\rm u}\,,
\end{equation}
with fractional uncertainty $1.3\times10^{-9}$. This is the most
accurate spectroscopic result for $m_{p}$ to date, and improves by
a factor 3.3 on the recent result \cite{Biesheuvel2016}.

In conclusion, with the newly introduced method, we increase the resolution
of rotational spectroscopy by a factor of 50.  Our experiment - theory
agreement between $f_{{\rm rot,exp}}$ and $f_{{\rm rot,theory}}$
is a direct proof that the present method permits rotational frequency
inaccuracy at least at the $9\times10^{^{-10}}$ level, without the
necessity for corrections. The reason for this are the method's resolution
and the favorable conditions in the UHV ion trap and not that the
HD\textsuperscript{+} test molecule is particularly \textit{in}sensitive
to perturbations.  Therefore, similar inaccuracy levels should be
achievable for other molecular species. 

The present work opens outstanding perspectives for precision physics.
With improvements in signal strength, we expect that the spectroscopic
resolution can be increased by at least one order, leading in addition
to a systematic uncertainty $<3\times10^{-11}$ (see Methods). This
will allow testing the \textit{ab initio} prediction of the rotational
frequency of the HD\textsuperscript{+} molecular ion with even higher
precision. The major contribution to $f_{{\rm rot,theory}}$, the
spin-averaged contribution, has already been evaluated in this work
with $1.4\times10^{-11}$ uncertainty and further progress on the
minor contributions, the hyperfine Hamiltonian coefficients, is expected
 \cite{Korobov2014,Korobov2014a,Korobov2016,Korobov2017a}. Such
an improved test will also provide a fairly direct test of theoretical
approaches used to compute the frequencies of the related molecular
ions ${\rm H}_{2}^{+}$ and ${\rm D}_{2}^{+}$ and of antiprotonic
helium. Eventually, the fundamental constants $R_{\infty},\,m_{e},\,m_{p},\,m_{d}$
will become measurable at the $10^{-11}$ level by \textit{molecular}
spectroscopy, which is a complementary approach to the current ones,
thus strengthening the overall consistency of this set of fundamental
constants. 

Thanks to the spectral resolution possible with the present method,
it will become possible to measure molecular properties which are
of small magnitude and thus are otherwise difficult access, e.g. level-dependent
magnetic moments and a.c. polarisabilities (light shifts). 

Our method also opens up vast possibilities in terms of accessible
molecular species. Given the mass ($m_{c}$) of singly-ionized laser-coolable
atomic ions, molecular ions with mass-to-charge ratio $m/q^{2}<m_{c}/e^{2}$
can be confined inside the atomic ion cluster, in the rotational Lamb-Dicke
regime. Employing, e.g., the high-mass ytterbium ion $^{171}\text{Yb}^{+}$
as coolant, many thousands of singly-ionized $(q=e)$ molecular species,
not counting isotopologues, have suitable mass. Many of these species
exhibit a simpler spin structure than the test case used here, which
can be advantageous in simplifying the spectrum and increasing state
populations. 

We note that radiation sources of appropriate spectral purity and
stability can be implemented not only using an H-maser as reference
but also more simply and accessibly using a GPS-steered high-performance
quartz oscillator. For most molecules, the fundamental rotational
transition frequency is smaller than $1\,$THz, which simplifies the
source further. In a complementary direction, the method could also
be applied to other rotational transitions $N>0\rightarrow N'=N+1$,
with correspondingly higher frequencies and thus potentially higher
fractional spectral resolution. Possibly, also two-photon \cite{Constantin2016}
and electric quadrupole rotational transitions could become accessible. 

Finally, this method should be applicable also to rotational stimulated
Raman transitions driven by co-propagating waves and to two-photon
vibrational transitions. The latter would be driven by counter-propagating
waves $f_{1}$, $f_{2}$ having a sufficiently small frequency difference
$|f_{1}-f_{2}|<c/2\pi\delta x$, ensuring the LDR. It is particularly
advantageous that in diatomics transitions exist for which $f_{1}$
and $f_{2}$ can be chosen to satisfy this condition but also to be
near-resonant with a dipole-allowed transition to an intermediate
rovibrational level, so that the two-photon transition rate is enhanced.

\begin{figure}[t]
\begin{centering}
\includegraphics[width=0.7\columnwidth]{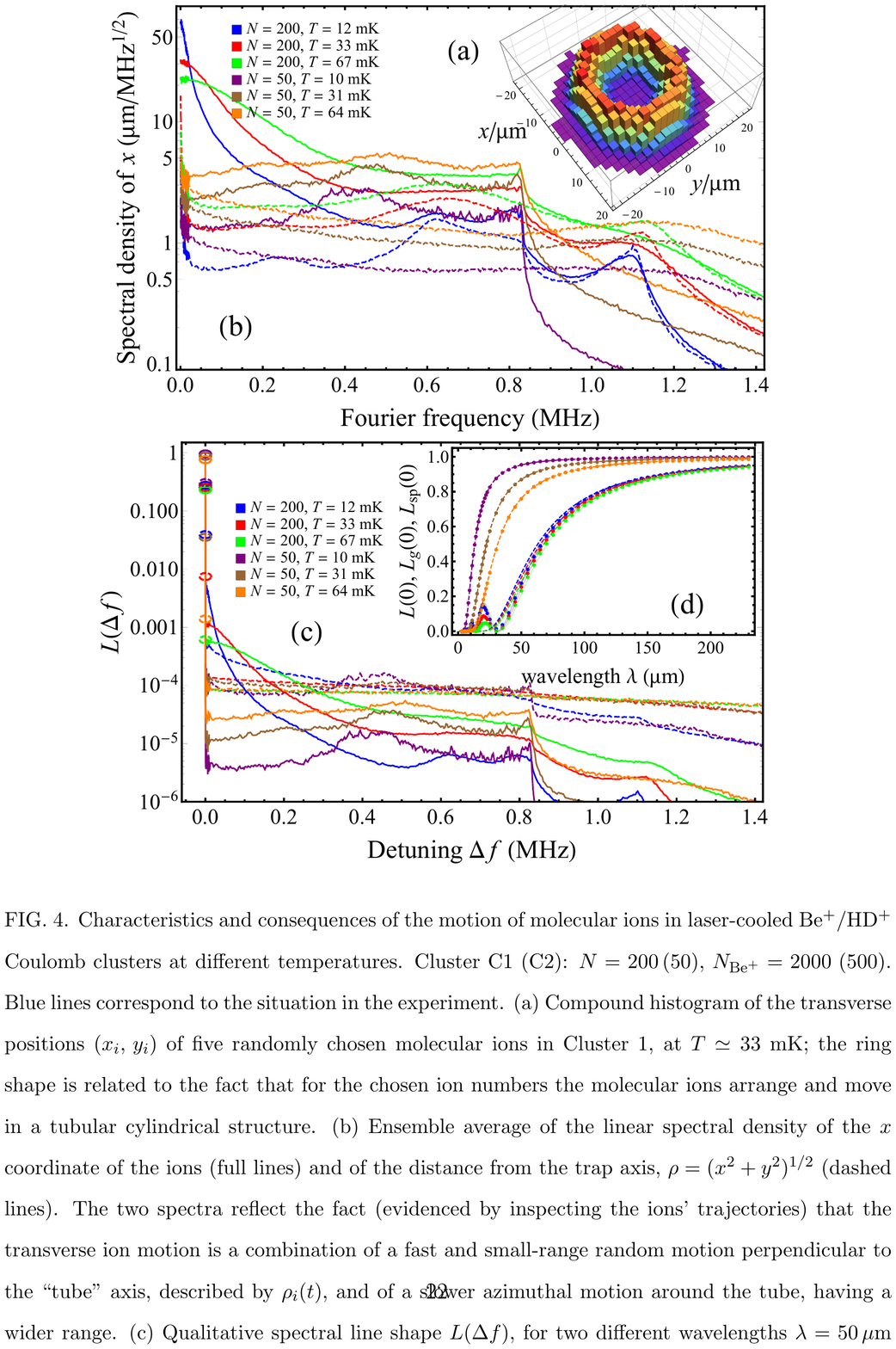}
\par\end{centering}
\caption{{\scriptsize{}\label{fig:time evolutions of coordinates-1}Characteristics
and consequences of the motion of molecular ions in laser-cooled Be}\protect\textsuperscript{+}{\scriptsize{}/HD}\protect\textsuperscript{+}{\scriptsize{}
Coulomb clusters at different temperatures. Cluster C1~(C2): $N=200\,(50),$
$N_{{\rm Be^{+}}}=2000\text{ }(500)$. Blue lines correspond to the
situation in the experiment. (a) Compound histogram of the transverse
positions $(x_{i},\,y_{i})$ of five randomly chosen molecular ions
in Cluster 1, at $T\simeq33$~mK; the ring shape is related to the
fact that for the chosen ion numbers the molecular ions arrange and
move in a tubular cylindrical structure. (b) Ensemble average of the
linear spectral density of the $x$ coordinate of the ions (full lines)
and of the distance from the trap axis, $\rho=(x^{2}+y^{2})^{1/2}$
(dashed lines). The two spectra reflect the fact (evidenced by inspecting
the ions' trajectories) that the transverse ion motion is a combination
of a fast and small-range random motion perpendicular to the ``tube''
axis, described by $\rho_{i}(t)$, and of a slower azimuthal motion
around the tube,  having a wider range. (c) Qualitative spectral
line shape $L(\Delta f)$, for two different wavelengths $\lambda=50\,\mu$m
(full), $10\,\mu$m (dashed). Only the positive detunings are shown.
(d) Strength of the Lamb-Dicke peak $(\Delta f=0)$. Points: exact
values $L(0)$; dashed and full lines: approximate expressions $L_{{\rm g}}(0),$
$L_{{\rm sp}}(0)$, respectively, defined in the text. Same colours
in (b) and (c) and (d) correspond to the same cluster type and temperature.
Transverse secular frequency of the molecular ions is 0.81~MHz; a
corresponding feature is seen in the spectra of $x$ and in the line
shapes $L$. }}
\end{figure}

\begin{figure}[t]
\begin{centering}
\includegraphics[width=0.9\columnwidth]{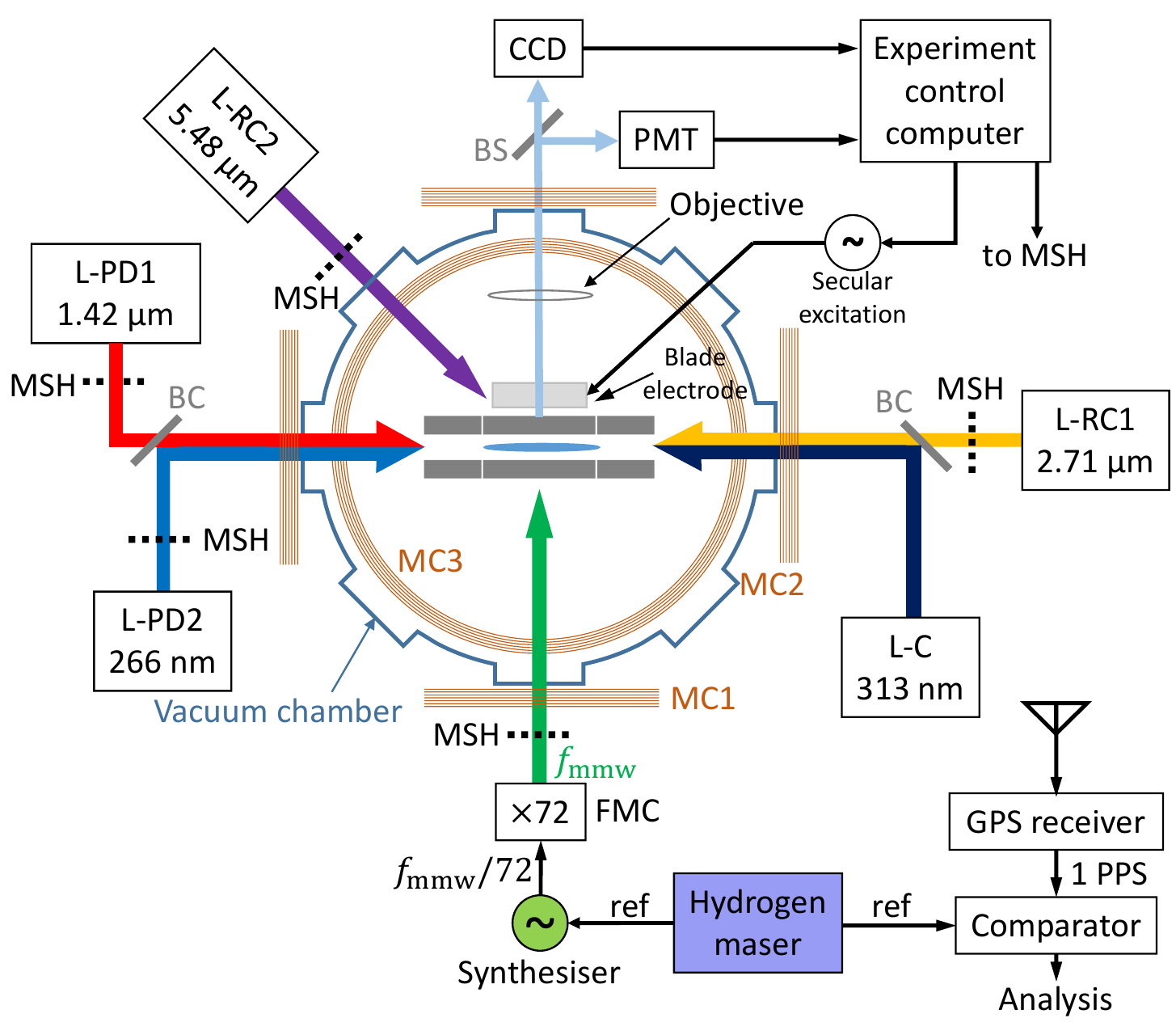}
\par\end{centering}
\caption{\label{fig:Simplified-schematic}Simplified schematic of the apparatus.
The THz radiation (green arrow) propagates along $x$, perpendicularly
to the ion cluster\textasciiacute s axis. BC: beam combiner, BS: beam
splitter, FMC: frequency multiplier chain, MSH: mechanical shutter,
L-PD: photo-dissociation lasers, L-RC: rotational cooling lasers,
L-C: Be\protect\textsuperscript{+} cooling laser, PMT: photo-multiplier
tube, MC: magnetic coils, CCD: CCD camera. Beryllium oven, electron
gun, and HD gas handling system are not shown.}
\end{figure}

\section{Methods}

\subsection{Simulation of ion dynamics in Coulomb clusters}

The spatial distribution of ions in a two-species Coulomb crystal
is well-known \cite{Hornekaer2001}. Given an elongated trap and electrode
geometry, and for typical magnitudes of the RF voltages and of the
end-cap voltages, a strongly prolate atomic ion cluster can result.
Sympathetically cooled molecular ions, if lighter than the atomic
coolant ions, and if in smaller number, are distributed in a string-like
or cylinder-like volume, centered on the trap axis ($z$). (Equal
charge states are assumed.) The boundary shapes of the two species'
spatial distributions are constant in time. 

Molecular dynamics (MD) simulations help to elucidate details, in
particular effects of temperature and dynamics. We simulated two ion
clusters. Cluster C1 contains $N_{{\rm Be^{+}}}=2000$~Be\textsuperscript{+}~ions
which sympathetically cool $N=200$~HD\textsuperscript{+} ions,
resulting in a tubular configuration for the latter. C1 is similar
to the experimentally produced clusters (Fig.~\ref{fig:Basic principle}).
A smaller cluster (C2) was modeled for comparison: with $N=50$, $N_{{\rm Be^{+}}}=500$
the molecular ions arrange as a string. The simulations are performed
in the pseudopotential approximation, since the micromotion of HD\textsuperscript{+}
ions only gives small corrections (here, the $q$-parameter is 0.15).
The simulations extended over 5~ms. The equilibrium secular temperature
$T$ of the molecular ions is determined by the assumed cooling and
heating rate parameters, which are varied in order to exhibit the
temperature dependence of the clusters' properties \cite{Zhang2007}.
A key feature of large clusters such as C1 is that even at the lowest
temperatures achievable experimentally by sympathetic cooling, $T\simeq10-30\,{\rm mK}$,
the ions' positions are not completely ``frozen''. Instead, the
ions diffuse through the cluster volume, with the diffusion speed
being a function of temperature \cite{Zhang2007}. 

The ion motion characteristics can be analyzed in detail. One characteristic
is the coordinate r.m.s. variation, defined as $\Delta\xi=\left\langle \Delta\xi_{i}\right\rangle $,
$\Delta\xi_{i}=\overline{\xi_{i}(t)^{2}}-\overline{\xi_{i}(t)}^{2}$,
where $i$ numbers the ion, $\xi=x,\,y,\,z$, the over-bar denotes
the time average and $\langle\ldots\rangle=N^{-1}\sum_{i=1}^{N}\ldots$
denotes the ensemble average. The values for C1 put in evidence transverse
confinement and diffusion features: whereas for motion along the trap
axis $\Delta z_{{\rm C1}}=(41,\,142,\,206)\,\mu{\rm m}$ over a time
interval of 5~ms, at $T=(12,\,{\rm 33,\,{\rm 67}})\,{\rm mK}$, respectively,
the transverse excursions are $\Delta x_{{\rm C1}}=\Delta y_{{\rm C1}}=(8.4,\,8.6,\,9.0)\,\mu{\rm m}$.
These latter values are significantly smaller than $\lambda_{{\rm rot,min}}/2\pi$.
In Fig.~\ref{fig:time evolutions of coordinates-1}~(a,~b) we show
some details of HD\textsuperscript{+} ion motion, the spectral density
of the $x$ coordinate and the time-averaged spatial distribution. 

We can obtain an estimate for the spectroscopy line shape $L$ induced
by the ions' motion, neglecting recoil and spontaneous emission effects,
by evaluating the ensemble-averaged Fourier spectrum of the spectroscopy
wave's electric field amplitude seen by a moving ion, $E({\bf r}_{i},\,t)\propto{\rm Re}\left[{\rm exp}(i\Phi({\bf r}_{i}(t))-i\omega t)\right]$,
where $\Phi({\bf r}_{i})={\bf k}\cdot{\bf r}_{i}$. For transverse
excitation along $x$, $\Phi({\bf r}_{i})=2\pi x_{i}/\lambda$. $\omega$
and $\lambda$ are the angular frequency and wavelength of the wave.
We define the line shape as $L(\Delta f)=\left\langle \left|\mathcal{F}_{\Delta f}\left\{ {\rm exp}(i\,2\pi x_{i}(t)/\lambda)\right\} \right|^{2}\right\rangle $,
where the argument $\Delta f$ of the Fourier transform corresponds
to the spectroscopic detuning from resonance. For wavelengths $\lambda\lesssim10\,\mu\text{m}$
the line shape is close to Gaussian, with the width $\Delta f_{{\rm FWMH}}$
determined by the classic expression proportional to $\sqrt{T}$,
see the dashed lines in Fig.~\ref{fig:time evolutions of coordinates-1}\,(c).
For wavelengths $\lambda\ge50\,\mu\text{m}$ a substantial delta function
peak $L(0)$ (Lamb-Dicke peak) develops at $\Delta f=0$. It is given
by $L(0)=\left\langle \left|\overline{{\rm exp}(i\,2\pi x_{i}(t)/\lambda)}\right|^{2}\right\rangle $
and is shown in Fig.~\ref{fig:time evolutions of coordinates-1}\,(d)
(points) as a function of wavelength for the two clusters. 

If the ions' coordinates $x_{i}(t)$ were Gaussian random variables,
$L(0)$ would simplify to $L_{{\rm g}}(0)=\left\langle {\rm exp}(-(2\pi\Delta x_{i}/\lambda)^{2})\right\rangle $
\cite{Cramer1970}. Since in clusters of the sizes as in C1, C2, most
ions behave similarly, $L_{{\rm g}}(0)\simeq{\rm exp}(-(2\pi\Delta x/\lambda)^{2})$.
This expression explicitly shows the wavelength dependence, and is
presented as dashed lines in the figure. The Gaussian assumption is
well satisfied when the number of molecules $N$ is reduced so far
(cluster C2) that they arrange approximately like a string along the
$z$-axis, since their $x$ - histogram is then Gaussian. For C2, $\Delta x_{{\rm C2}}=\Delta y_{{\rm C2}}\simeq(1.7,\,3.0,\,4.1)\,\mu\text{m}$
at $T=(12,\,{\rm 33,\,{\rm 67}})\,{\rm mK}$ increase approximately
with the square root of the temperature, and thus more strongly in
relative terms than for C1. This leads to a more pronounced variation
of $L(0)$ with temperature than for C1, see Fig.~\ref{fig:time evolutions of coordinates-1}\,(d).
The small values $\Delta x_{{\rm C2}},\,\Delta y_{{\rm C2}}$ lead
to a substantial $L(0)\simeq0.3$ already for $\lambda\simeq10\,\mu\text{m}$
when $T=13\,$mK. 

For the C1 cluster the Gaussian assumption is not correct (see histogram
in Fig.~\ref{fig:time evolutions of coordinates-1}\,(a)), therefore
deviations between $L_{{\rm g}}(0)$ and $L(0)$ are visible in Fig.~\ref{fig:time evolutions of coordinates-1}\,(d)
for intermediate and small wavelengths. Heuristically, we find that
the expression $L_{{\rm sp}}(0)=J_{0}(2\pi\sqrt{2}\Delta x/\lambda)^{2}$,
the result for a single ion harmonically oscillating along the $x$
axis \cite{Berkeland1998a}, provides a better description of the
exact values because of the similarity in the probability distributions
of the $x$ coordinate values.

In summary, the simulations indicate that even for only moderately
cold ensembles ($T\simeq70\,$mK) and for a relatively small value
$\lambda_{{\rm rot,min}}=228\,\mu{\rm \text{m}}$ ($f_{{\rm rot,max}}=1.3$~THz)
strong signatures of Lamb-Dicke confinement in rotational spectroscopy
with transverse incidence can be expected. For string-like ion clusters
a similar signature might be possible also for \textit{axial} irradiation,
but only at the lowest temperatures $(T\simeq10\,{\rm mK})$, when
ion position changes are infrequent. This case will be the subject
of future experimental studies.

\subsection{Experimental Apparatus}

The ion trap apparatus used in the present work, Fig.~\ref{fig:Simplified-schematic}
is based on the device used previously \cite{Blythe2005,Schneider2010,Bressel2012,Shen2012},
and upgraded in several respects: (i) fully computer-controlled operation,
(ii) accurate magnetic field control via solenoid pairs, (iii) improved
frequency stabilization of the rotational cooling lasers L-RC1 and
L-RC2, (iv) improved THz source frequency control.

The vacuum chamber housing the ion trap provides ultra-high vacuum
conditions ($3\times10^{-11}$~mbar), minimizing spectral broadening
and shifts due to ion - background gas collisions. Electron impact
ionization inside the ion trap volume is used to generate Beryllium
ions from Be atoms emitted by a hot filament and HD\textsuperscript{+}
ions from HD gas injected into the chamber. 313~nm radiation laser
cools the Beryllium ions which sympathetically cool the HD\textsuperscript{+}
ions, resulting in a structured ion cluster as shown in Fig.~\ref{fig:Basic principle}.
Subsequently, radiation fields at $5.48\,\mu$m and $2.71\,\mu$m
perform rotational cooling. The 1.3~THz radiation for rotational
spectroscopy propagates at 90~degrees with respect to the ion trap
axis and thus the ion cluster axis, in order to provide the LDR. The
beam radius at the center of the trap is approximately 1~mm. A CCD
camera images the cluster's spatial structure, allowing for a direct
observation of the spatial confinement and also comparison with molecular
dynamics simulations. The photo-multiplier tube detects the fluorescence
of the $\text{Be}^{+}$ ions, and also provides the signal for the
spectroscopy.

\begin{figure}[t]
\noindent \begin{centering}
\includegraphics[width=0.9\columnwidth]{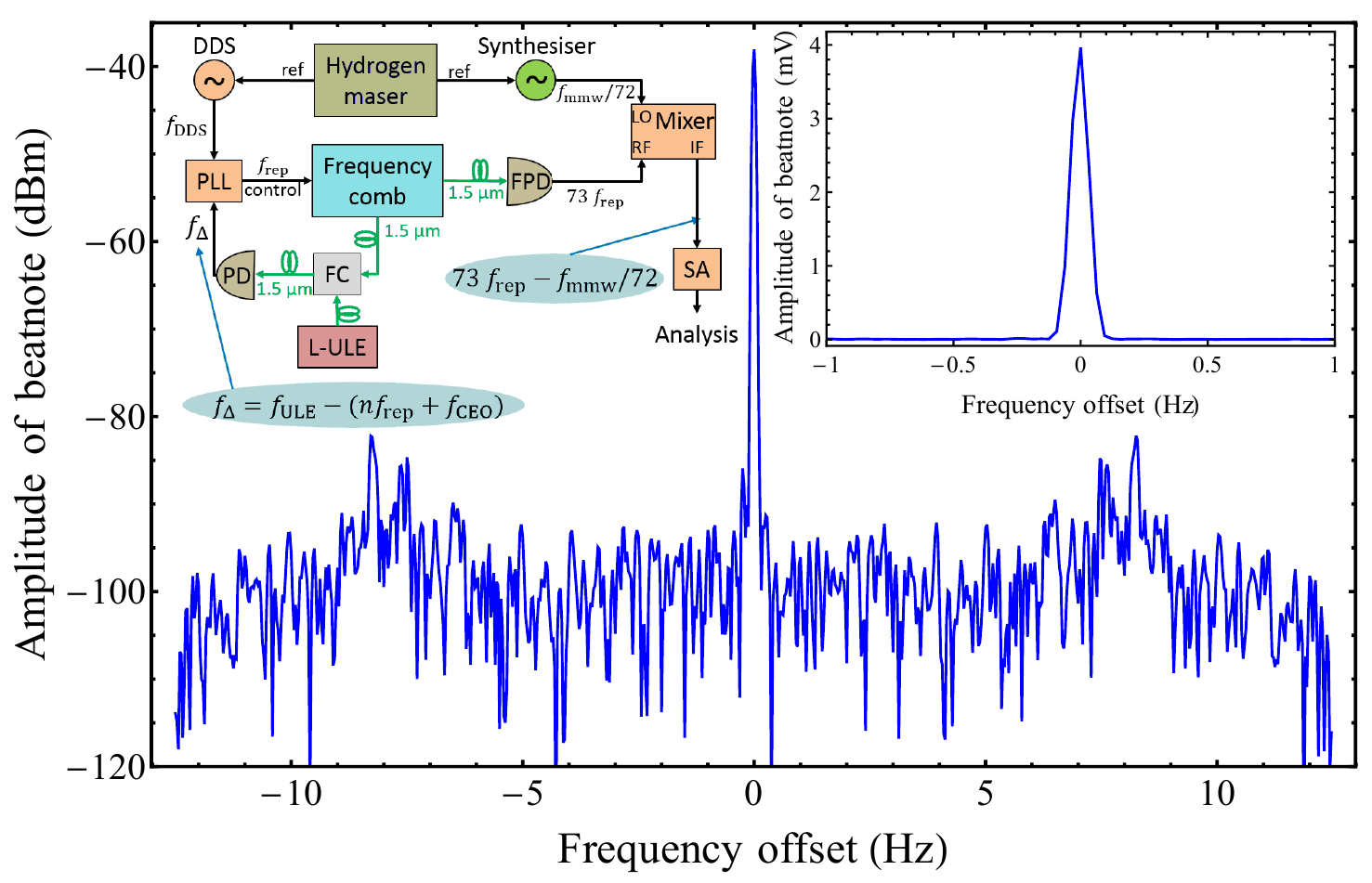}
\par\end{centering}
\caption{\label{fig:schematic_agilent_linewidth}THz line width characterization.
Main plot: power spectrum of the signal oscillating at the difference
between an intermediate frequency close to $f_{{\rm mmw}/72}$ (synthesizer)
and $f_{{\rm rep}\times73}$, the 73\protect\textsuperscript{rd}
harmonic of the frequency comb's repetition rate. Full-width half-maximum
is $86\,\text{mHz}$, with the resolution bandwidth being 31.25~mHz.
Right inset: same as the main plot but the amplitude is in mV. Left
inset: Schematic of the line-width measurement setup. FC: Fiber combiner,
L-ULE: continuous-wave $1.5~\mu$m semiconductor laser stabilized
to an ultra-low expansion (ULE) glass cavity, PD: photo-detector,
FPD: high-bandwidth photo-detector, SA: spectrum analyzer, PLL: phase
lock feedback loop, DDS: direct digital synthesizer.}
\end{figure}

The THz system consists of a hydrogen maser, whose frequency, $f_{{\rm H}}\simeq1.4\,{\rm GHz}$,
is down-converted to 10~MHz and then used as reference for a microwave
synthesizer operating at $f_{{\rm mmw/72}}=18.262$~GHz. Its output
is converted in a $\times72$ multiplier/amplifier chain to $f_{{\rm rot}}=$1.31~THz
\cite{Schiller2009a}. We verified the high spectral purity of the
THz wave at the intermediate frequency 18.251~GHz, close to $f_{{\rm mmw/72}}$.
The setup used is shown in Fig.~\ref{fig:schematic_agilent_linewidth}~(inset).
The repetition rate $f_{{\rm rep}}\simeq250\,\text{MHz}$ of a fiber-based
optical frequency comb is stabilized by phase-locking the beat frequency
between an ultra-stable reference laser at $1.5~\mu$m (L-ULE, with
optical frequency $f_{{\rm ULE}}$ \cite{Wiens2016}) and a nearby
comb mode (mode number $n\simeq7.8\times10^{5}$) to a (maser-referenced)
DDS set at $f_{\text{DDS}}=50$ MHz, by controlling the repetition
rate. The comb carrier-envelope offset frequency $f_{{\rm CEO}}$
is independently stabilized to the H-maser. The absolute frequency
stability of the reference laser is therefore transferred to the repetition
rate, $f_{{\rm rep}}=(f_{\text{ULE}}-f_{{\rm DDS}}-f_{\text{CEO}})/n$.
Comb radiation at $1.5\,\mu\text{m}$ is detected by a fast photo-detector,
and an RF signal arising from the 73\textsuperscript{rd} harmonic
of the repetition rate, $f_{{\rm rep}\times73}$, is generated. This
harmonic is chosen since it can be set close to $f_{{\rm mmw/72}}$.
The difference frequency between $f_{{\rm rep}\times73}$ and $f_{{\rm mmw}/72}$
is generated using a mixer and is analyzed using an FFT-based spectrum
analyzer. Fig.~\ref{fig:schematic_agilent_linewidth} shows the
line width of the difference frequency signal, which is $86\,\text{mHz}$.
Based on this value and previous characterizations \cite{Schiller2009a},
we estimate a line width $<10\,{\rm Hz}$ at 1.3~THz. We can also
infer the long-term frequency stability of the THz wave, by comparing
the maser frequency to a GPS-derived 1~PPS signal and to a cryogenic
silicon optical resonator \cite{Wiens2014}. The residual instability
is negligible ($<1\times10^{-13}$  for integration times $\tau>10\,{\rm s}$)
and so is drift.

\noindent 

\subsection{Laser rotational cooling}

 In the present experiment, when molecular ions are generated and
trapped and reach thermal equilibrium, the $(v=0,\,N=0)$ level's
population is $\simeq$10\%, and thus only $\simeq$1\% is in a single
state $|g,\,J_{z}\rangle$. Therefore, before spectroscopy, we apply
rotational laser cooling \cite{Schneider2010} (lasers L-RC1, L-RC2
in Fig.~\ref{fig:Simplified-schematic}), to increase significantly
the population in $(v=0,\,N=0)$, a procedure which also increases
the population in the Zeeman states $|g,\,J_{z}\rangle$ to a level
which can be observed. 

We apply two laser fields which drive the fundamental vibrational
transition $(v=0,\,N=2)\rightarrow(v'=1,\,N'=1)$, and the overtone
vibrational transition $(v=0,\,N=1)\rightarrow(v'=2,\,N'=0)$. Because
of quantum mechanical selection rules, repeated absorption - spontaneous
emission cycles transfer the majority of the ion population into the
ground state $(v=0,\,N=0)$. 

A distributed feedback (DFB) laser at 2.71~$\mu\text{m}$ is used
to excite the overtone vibrational transition (L-RC1 in Fig.~\ref{fig:Simplified-schematic}).
The laser is frequency stabilized to a $\text{CO}_{2}$ gas transition
using an offset locking technique \cite{Nevsky2013} to bridge the
frequency gap between the $\text{CO}_{2}$ transition and the $\text{HD}^{+}$
transition frequency. We use a quantum cascade laser (QCL) to drive
the fundamental transition at 5.48~$\mu\text{m}$ (L-RC2 in Fig.~\ref{fig:Simplified-schematic}),
whose frequency is stabilized to the side of a fringe of an $\text{NH}_{3}$
molecular transition.

\subsection{Experimental sequence.}

The preparation and spectroscopy sequence consists of the following
steps: (1) HD\textsuperscript{+} generation by electron impact ionization;
(2) impurity ion removal procedures; (3) rotational cooling for $t_{{\rm rc}}=35\,\text{s}$;
(4) secular excitation for $t_{{\rm se}}=3\,\text{s}$, with rotational
cooling lasers on; (5) during $t_{{\rm det}}=3\,\text{s}$, the rotational
cooling lasers are blocked, while THz radiation, and the $1.42\,\mu$m
and 266~nm waves for REMPD are on; (6) secular excitation during
$t_{{\rm se}}$. The signal is obtained from the difference in Be\textsuperscript{+}
fluorescence recorded during intervals (4) and (6).

The magnetic field is set to 1.1~G along the trap axis $z$, except
during step (5), where it is set to a smaller value $B$, sufficient
to produce Zeeman splitting of most components of the studied transition.
The sequence is repeated a number of times sufficient to obtain a
reasonable signal-to-noise ratio. Be\textsuperscript{+} ions are
reloaded every 100 to 150 sequences.

\subsection{Rotational transition}

The fundamental rotational transition of the molecular ion used here
has an important spin structure because of the nonzero electron and
nuclear magnetic moments \cite{Bakalov2006}. While in the ground
rovibrational level $(v=0,\,N=0)$ they give rise to 4~spin states
(having total angular momentum $J=0,\,1,\,2)$, in the first excited
rotational level $(v=0,\,N'=1)$ the number increases to 10 spin states
(having $J'=0,\,1,\,2,\,3)$, because of the additional presence of
rotational angular momentum $N'\ne0$. Fig.~\ref{fig:Energy diagram-1}
shows the two spin states relevant to the spectroscopy of this work
(for a complete diagram, see Fig.~2 in~\cite{Shen2012}). A controllable
magnetic field lifts the total angular momentum projection ($J_{z}$)
degeneracies, except for the two transitions ${\rm T}_{\pm}$. We
address transitions between two individual Zeeman components $|g,\,J_{z}\rangle$
of the $J=2$ spin state of the $N=0$ - level, and two components
$|e,\,J_{z}'\rangle$ of the $J'=3$ spin state of the $N'=1$ - level.
The transitions T\textsubscript{+}: $J_{z}=2\rightarrow J_{z}'=3$
and ${\rm T}_{-}$: $J_{z}=-2\rightarrow J_{z}=-3$, shown by green
arrows in Fig.~\ref{fig:Energy diagram-1}, are transitions between
stretched states (states of maximal total angular momentum $J$ and
maximal projection $J_{z}$) and have been previously identified as
exhibiting particularly low Zeeman shifts \cite{Bakalov2011,Bakalov2014}.

\subsection{\textit{Ab initio} theory of the HD\protect\textsuperscript{+} rotational
transition}

The transition frequency $f_{{\rm rot,theory}}$ involves two contributions.
The first is the spin-averaged frequency $f_{{\rm spin-avg}}$. We
have computed it with very high precision using the same technique
as in \cite{Korobov2017a}, obtaining 
\[
f_{{\rm spin-avg}}=1\,314\,925.752\,627(18)\mbox{ MHz},
\]
with the relative uncertainty $1.4\times10^{-11}$. The second contribution
is the hyperfine shift $f_{{\rm spin}}$ due to spin interactions
(including coupling with the total orbital angular momentum), $f_{{\rm spin}}=[E_{{\rm spin}}(|u\rangle)-E_{{\rm spin}}(|l\rangle]/h$.
Here $E_{{\rm spin}}(|l\rangle)$ and $E_{{\rm spin}}(|u\rangle)$
are the hyperfine spin energy shifts for the individual lower and
upper states. So far, they have been calculated within the Breit-Pauli
approximation only \cite{Bakalov2006}. The spin energy shifts of
the stretched states $|g,\,J_{z}=\pm2\rangle$ and $|e,\,J_{z}'=\pm3\rangle$
in a finite magnetic field can be readily computed using the algebraic
expression (Eq.~(6) in \cite{Bakalov2011}),

\begin{align*}
E_{{\rm spin}}(v,\,N,\,J=L+2,\,J_{z}=\pm J,\,B)/h=\,\,\,\,\,\,\,\,\,\,\,\,\,\,\,\,\,\,\,\,\,\\
\pm(2E_{10}N+E_{11}+2E_{12}+E_{13})\,B/2+E_{4}/4+E_{5}/2\\
+(E_{1}+E_{2}+2E_{3}+E_{6}+2E_{7}+2E_{8}+E_{9})\,N/2\\
-(2E_{6}+4E_{7}+4E_{8}+2E_{9})\,N^{2}/2\,,
\end{align*}
where $E_{i}=E_{i}(v,\,N)$ are the coefficients of the effective
spin Hamiltonian \cite{Bakalov2006,Bakalov2011}. Since the $J=2\rightarrow J'=3$
transition keeps the spin function untouched, the hyperfine shift
$f_{{\rm spin}}$ is most sensitive to the electron spin-orbit interaction
$E_{1}(v=0,\,N=1)({\bf s_{e}{\bf \cdot}}{\bf N})$, where $E_{1}(v=0,\,N=1)\simeq32\,{\rm MHz}$
\cite{Bakalov2006}. The spin-spin interactions, proportional to $E_{4}$
and $E_{5}$, which are known with much higher precision than the
other coefficients \cite{Korobov2016}, give a much smaller contribution
to $f_{{\rm spin}}$, due to the similarity of the spin wave functions
that leads to a substantial cancellation. The hyperfine shift is

\begin{equation}
f_{{\rm spin}}=10.0747(10){\rm \text{ }MHz}.
\end{equation}

 Summing both above contributions we obtain eq.~(\ref{eq:f_rot,theory}). 

In order to improve the 1-kHz theoretical uncertainty of $f_{{\rm spin}}$
it will be necessary to calculate the higher-order corrections to
the coefficient $E_{1}$. This particular sensitivity of the observed
transition on the accuracy of the effective spin Hamiltonian is characteristic
for rotational transitions. For vibrational transitions, the much
larger ratio of $f_{{\rm spin-avg}}$ to $f_{{\rm spin}}$ reduces
the sensitivity significantly. This is the case for the most intensive
hyperfine components of the transitions (for which $|f_{{\rm spin}}|<100\,{\rm MHz}$),
and the Breit-Pauli approximation is then in most cases sufficient
to guarantee a fractional uncertainty of the total transition frequency
comparable to the fractional uncertainty of the theoretical spin-averaged
frequency. 

It is important to note that the theoretical considerations of the
spin-averaged and the hyperfine spin shift contributions to $f_{{\rm rot,theory}}$
have recently been confirmed through a comparison, respectively, with
one particular rovibrational transition frequency of HD\textsuperscript{+}
at the 1.1~ppb level \cite{Biesheuvel2016} and with several spin
transition frequencies (i.e. radio-frequency transitions) of H$_{2}^{+}$
at the ppm level \cite{Korobov2016}.

\subsection{Systematic shifts}

We have previously computed  most relevant sensitivities of the rotational
transition to external perturbations: Zeeman shift \cite{Bakalov2011},
electric quadrupole shift \cite{Bakalov2014}, d.c. Stark shift, black-body
radiation shift, and spin-state dependence of the d.c. Stark and light
shift \cite{Schiller2014a}. The quadratic Zeeman shift is zero for
the transitions ${\rm T}_{\pm}$. 

For the experimental parameters of our trap, the cluster shape, and
the moderate intensities of the radiation fields, the shifts are all
small, except for the Zeeman shift. Since the orientation of the quantization
axis and the populations of the lower Zeeman states are unknown, we
assign an uncertainty equal to half the Zeeman splitting of T\textsubscript{+}
and ${\rm T}_{-}$, 0.22~kHz. The d.c. Stark shift has been computed
taking into account ion trajectories and Coulomb fields, also allowing
for d.c. offset potentials, and is less than 10~Hz. Collision shifts
related to background gas are negligible due to the UHV conditions.
One finite shift is the light shift induced by the 266~nm dissociation
laser (35~mW power). In order to compute it, we have performed a
precision calculation of the a.c. polarisabilities of the lower and
upper rotational levels using the procedure described in \cite{Schiller2014a}.
We obtained the scalar ($s$) and tensor ($t$) polarisabilities $\alpha_{s}(v=0,\,N=0\,(1),\,\lambda=266.0\,{\rm nm})=3.677\,(3.687)$,
$\alpha_{t}(v=0,\,N=0\,(1),\,\lambda=266.0\,{\rm nm})=0\,(-1.044)$,
in atomic units. For the considered transitions, the shift is 7~(40)~Hz
In total, we obtain a frequency correction of 0.0(3)~kHz. 

In future studies, any light shifts from the dissociation laser and
from the $1.4\,\mu$m laser could be avoided by applying these lasers
only after the rotational excitation. If the Zeeman pair were split
and resolved, the Zeeman shift uncertainty should be reduced at least
10-fold. This would then allow a total systematic uncertainty of $<3\times10^{-11}$.

\smallskip{}
\begin{acknowledgments}
This work has been partially funded by DFG project Schi~431/21-1.
We thank U.~Rosowski for important assistance with the frequency
comb, A.~Nevsky for assistance with a laser system, E.~Wiens for
characterizing H-maser instability, R.~Gusek and P.~Dutkiewicz for
electronics development, J.~Scheuer and M.~Melzer for assistance,
and S.~Schlemmer (Universität zu Köln) for equipment loans. We thank
K.~Brown (Georgia Institute of Technology) for useful discussions
and suggestions.

Corresponding author, step.schiller@hhu.de

\textbf{\smallskip{}
Contributions}

S.A. and M.G.H. developed the apparatus and performed the experiments,
S.A., M.G.H., and S.S. analyzed the data, S.A., S.S. and V.I.K. performed
theoretical calculations, S.S. conceived the study, supervised the
work and wrote the paper.
\end{acknowledgments}

\textbf{\smallskip{}
Competing financial interests.}

The authors declare no competing financial interests.

\bibliographystyle{elsarticle-num}
\phantomsection

\cleardoublepage{}
\end{document}